\documentclass[conference]{IEEEtran}  % Comment this line out
\IEEEoverridecommandlockouts                              % This command is only
\usepackage[utf8]{inputenc}
\usepackage[T1]{fontenc}
\usepackage{amsfonts}

\usepackage{graphicx}
\graphicspath{ {./figures/} }

\usepackage[
backend=biber,
style=numeric,
citestyle=numeric,
sorting=none
]{biblatex}
\usepackage{url}

\usepackage{breakurl}
\usepackage[breaklinks]{hyperref}

\title{
Data Capsule: A Self-Contained Data Model as an Access Policy Enforcement Strategy
}

\author{

\IEEEauthorblockN{
Reza Soltani\IEEEauthorrefmark{1}, 
Uyen Trang Nguyen\IEEEauthorrefmark{2} and 
Aijun An\IEEEauthorrefmark{3}}
\IEEEauthorblockA{Lassonde School of Engineering\\
York University\\
Toronto, Canada\\
Email: \IEEEauthorrefmark{1}rts@cse.yorku.ca,
\IEEEauthorrefmark{2}utn@cse.yorku.ca,
\IEEEauthorrefmark{3}aan@cse.yorku.ca
}}

\begin{document}

\maketitle
\thispagestyle{empty}
\pagestyle{empty}

%%%%%%%%%%%%%%%%%%%%%%%%%%%%%%%%%%%%%%%%%%%%%%%%%%%%%%%%%%%%%%%%%%%%%%%%%%%%%%%%
\begin{abstract}

In this paper, we introduce a data capsule model, a self-contained and self-enforcing data container based on emerging self-sovereign identity standards, blockchain, and attribute-based encryption. A data capsule allows for a transparent, privacy-respecting, and secure exchange of personal data, enabling a progressive trust scheme in a semi-trusted environment. Each data capsule is bundled with its own access policy structure and verifiable data, drastically reducing the number of interactions needed among the user, the service providers, and data custodians. Moreover, by relying on the decentralized nature of blockchain and attribute-based encryption our proposed model ensures the access policies published by service providers are public, transparent, and strictly followed.

\end{abstract}

%%%%%%%%%%%%%%%%%%%%%%%%%%%%%%%%%%%%%%%%%%%%%%%%%%%%%%%%%%%%%%%%%%%%%%%%%%%%%%%%

\section{INTRODUCTION}

With the advent of online interactions and data sharing, there is a growing trend for the digital exchange of personally identifiable information (PII).The steady increase in security incidents, data breaches, and unauthorized surveillance compromising users' privacy raises questions around the potentials for improving the existing identity and access management models used today that place service provider in the center of their ecosystem. In such models, service providers collect and control a massive amounts of personal data with none to very limited checks and balances in place.

While new privacy frameworks have emerged in recent years, there is still lack of transparency over the degree by which these frameworks are followed \cite{techcruch_privacy_2020}. Regulations such as GDPR\cite{gdpr_online} and CCPA\cite{ccpa} provide a framework around protecting personal data, yet it is still difficult to identity breaches and data misuse before they become public. We believe reliance on cryptographic protocols and decentralized identity to support granular access control over verifiable data to enforce data minimization and progressive trust is a possible solution to these challenges. 

Traditional access control mechanisms that rely on access control lists place the onus on the service providers as the key entities responsible for evaluation and enforcement of access policies. This could inadvertently or intentionally lead to unauthorized data misuse or disclosure.  A well known method to protect sensitive data stored by third party service providers is encryption. When encrypted data stores are compromised, the amount of data loss is generally less \cite{miller2011encryption}. However, there are many shortcomings with blank encryption of data including limited capability for granular and selective data disclosure and scalability challenges around  key management. 

In order to enable users with granular control over their data, a data model based on attribute-based encryption (ABE) and Self-Sovereign Identity model (SSI) is proposed.  In this model, the user produces a \textit{Data Capsule}. A data capsule contains the relevant access policy bundled with the encrypted data the user intends to share. Only the service providers with the attributes that satisfy the access policy are able to decrypt the data. 

\subsection{Our Contribution}
In order to equip users with granular control over their PII, we propose the Data Capsule model. Data Capsule is a privacy-preserving  data exchange model based on  attribute-based encryption, blockchain technology, Decentralized Identifiers (DID) and Verifiable Credentials (VC). The advantages of the Data Capsule model are: 
\begin{itemize}

\item Use of blockchain to support a decentralized trust model, transparent access policies, and enforcement of data integrity and non-repudiation. 
\item Introducing the experimental specification for encrypted Verifiable Credentials and Verifiable Presentations
\item Placement of Verifiable Presentations along with access control structure and keys associated to a specific entity within a self-contained Data Capsule
\item To the best of our knowledge this is the first research work that utilizes Hyperledger Indy, Verifiable Credentials and DID for the purpose of data exchange using ABE.
\end{itemize}

\subsection{Case Study}

In the context of taxation system, every year a government agency in charge of collecting taxes obtains various types of personal information from its citizens. While the personal information such as first and last name may be used immediately by the agency upon submission, the financial information are typically used at a later time by another department. In this scenario there is an opportunity for data minimization and progressive disclosure of data.  

GDPR Rec.39; Art.5(1)(c) \cite{gdpr_online} states that “The personal data should be adequate, relevant and limited to what is necessary for the purposes for which they are processed. This requires, in particular, ensuring that the period for which the personal data are stored is limited to a strict minimum.”.

The proposed model can address this requirement by allowing the users to construct their online form submission as multiple DCs. The first capsule contains the encrypted version of the personal information which will be consumed immediately by the service provider, while the second contains the encrypted financial data which will be accessible at a specific predetermined time for specific recipients.    

In the absence of DC, if a citizen wishes to encrypt their tax forms, they must either encrypt the entire data and provide the master key to the agency, thus letting them gain access to the entire data, or alternatively, the citizen (or a data custodian) must perform data decryption on ad hoc bases for every data request. In the latter option the secret key remains under the control of the citizen, but the citizen must be involved in every data transaction.

\section{Building Blocks}
In this section we describe some of the building blocks of the DC model. 
 
\subsection{Decentralized Identifiers}
Developed by World Wide Web Consortium (W3C),  Decentralized Identifiers  \cite{noauthor_decentralized_nodate} are
a key component of the SSI architecture. The DID specification defines a globally unique and cryptographic
identifier scheme. DIDs can be managed using a decentralized infrastructure such as blockchain. The ownership of a DID can be proven by a challenge-response scheme.

\subsection{Verifiable Credentials}

The \textit{Verifiable Credentials} \cite{noauthor_verifiable_nodate} is a data model specification developed by W3C Verifiable Credentials working group. The VC model describes an
interoperable data structure suitable for representing cryptographically
verifiable and tamper-proof claims.  Verifiable Presentation (VP) describes a data structure derived from one or more verifiable credentials and shared with a specific verifier. A VP is issued by one or more issuer.

The VC model supports the \textit{Terms of Use} property which can be added by an issuer of a VC or a holder of a VC to communicate the terms under which a VC was issued. However, this approach fails to support any cryptographic mechanism to enforce the terms of use policy. Rather it relies on the judgement and the capabilities of the VC parties involved, to identify, comply or report any violations of the terms. 

\subsection{Hyperledger Indy}

Hyperledger Indy \cite{noauthor_hyperledgerindy_nodate} is a
public permissioned DLT platform built to support SSI. 
Hyperledger Indy supports the storage and mapping of DID to DID Descriptor Object (DDO), as well as the storage of various public identity related records on the ledger including public claims and key revocation entries. While our proposed model can be designed to work on other decentralised platforms, We've selected Hyperledger Indy as the platform of choice due to it's full support for SSI principles, its growing and active community, and its adherence to interoperability and privacy by design principles. 

\subsection{Attribute Based Encryption}
Attribute based encryption provides a cryptographic approach to support fine-grained access control over data. ABE facilitates the validation of an access policy over a set of attributes. Only the entities whose attributes satisfy the access policy set by the encryptor of the data, are able to decrypt the ciphertext. 
 
 There are two common forms of ABE, namely Ciphertext-Policy ABE  (CP-ABE) and Key-Policy ABE (KP-ABE). In \cite{goyal2006attribute} Goyal et al. provide a formal definition of CP-ABE and KP-ABE. In KP-ABE, the cryptographic key is associated with the access policy, and the ciphertext is associated with a set of attributes.  Conversely, in CP-ABE the key is associated with a set of attributes and the ciphertext is associated with the access policy. The choice of CP-ABE or KP-ABE depends on the application.  B. Waters describes the security model of CP-ABE scheme in \cite{waters2011ciphertext}.

\section{Underlying Algorithms}
The ABE schemes used in our model are designed based on the bilinear pairing and LSSS \cite{goyal2006attribute}.

The proposed model is based on attribute-based access control (ABAC)  \cite{rotondi2012managing} model in which the access policy is a Boolean formula over a set of attributes. 
Based on \cite{beimel1996secure} our access structure is defined as following. Let ${\{P_1, P_2,...,P_n\}}$ be a set of parties. A collection $ \mathbb{A} \subseteq 2^{\{P_1, P_2, ..., P_n\}}$ is a
monotone if $ \forall B, C : if B \in \mathbb{A} $ and $ B \subseteq C$ then $ C \in \mathbb{A}$. An access structure is a collection $\mathbb{A}$ of non-empty subsets of $\{P_1, P_2,...P_n\}$, 
i.e. $\mathbb{A} \subseteq 2^{P_1, P_2, ..., P_n} \textbackslash \{ \emptyset \} $. The sets in $\mathbb{A}$ are called the authorized sets, and the sets not in $\mathbb{A}$ are called the unauthorized set. 
In the context of proposed model, the access policies are restricted to monotone access structures, and $\mathbb{A} $ contains the authorized sets of attributes.  We take advantage of LSSS as defined in \cite{beimel1996secure} and \cite{waters2011ciphertext}. A secret-sharing scheme $\Pi$ over a set of parties $\mathbb{P}$ is called linear over $\mathbb{Z}_p$ if: 

\begin{enumerate}
    \item The shares for each party form a vector over $\mathbb{Z}_p$
    \item There exist a matrix $\mathbb{M}$ with $l$ rows and $n$ columns called the share-generating matrix for $\Pi$. For all $i = 1,.., l$, the $i'th$ row of $M$, we let the function $p$ define the party labeling row $i$ as $p(i)$. When we consider the column vector $v = (s, r_2, .., r_n)$, where $ s \in \mathbb{Z}_p$ is the secret to be shared, and $r_2,..,r_n \in \mathbb{Z}_p$ are randomly chosen, then $Mv$ is the vector of $l$ shares of the secret $s$ according to  $\Pi$. The share $(Mv)_i$ belongs to party $p(i)$.         
\end{enumerate}

Using standard techniques it is possible to convert any monotonic Boolean formula into an LSSS representation \cite{beimel1996secure}. An access tree  of $l$ nodes will result in an LSSS matrix of $l$ rows. 

  Finally, Bilinear pairing is defined as follows. Let $ \mathbb{G}_1, \mathbb{G}_2 $ be two multiplicative groups of prime order $g$ and let $g_1$ and $g_2$ be generators of $\mathbb{G}_1$ and $\mathbb{G}_2$, respectively. Then let us denote a bilinear map $ e : \mathbb{G}_1 \times \mathbb{G}_2 \to	 \mathbb{G}_T $. The map has the following three properties:
 \begin{enumerate}

  \item Bilinearity: $ \forall x \in \mathbb{G}_1, \forall y \in \mathbb{G}_2 $, and $ a,b \in \mathbb{Z}_q $, there is $ e(x_a, y_b) = e(x, y)^{ab}$

  \item Non-degeneracy: For $ \forall x \in \mathbb{G}_1, \forall y \in \mathbb{G}_2 $, there is $ e(x, y) \ne 1 $ 
  
  \item Computability: $ e $ is an efficient computation.

\end{enumerate}

\section{Proposed Model}

There are three main actors within our model.  A data owner (DO) is a user possessing the personal data. A service provider (SP) is the entity consuming the personal data. Finally, a data registry such as a blockchain framework is used to enable the decentralized trust model. The registry is used to publish decentralized identifiers and access policies. 

\subsection{Definition of the Model}

The structure of DC consists of three components. The metadata component holds the DC version and supported algorithms and blockchains. The second component consists of encrypted verifiable presentations and the access policy in the encrypted format. The third component holds the cryptographic key for the intended receivers of the capsule. As mentioned, the key contains in itself a set of attributes.
The model is composed of the following algorithms: 
\begin{itemize}
   
   \item $Setup(1^\lambda) \rightarrow PK, MSK$: This algorithm accepts the ABE security parameters as input, and outputs the public parameters $PK$ of the system and the master secret key $MSK$ for the data owner. 
   
   \item $KeyGen(PK, MSK, I, S) \rightarrow k$: This algorithm consists of three steps:
   \begin{enumerate}
       \item $Lookup(I) \rightarrow PK_I$ or $\bot$: This algorithm perform a registry lookup to obtain the public key of a given DID belonging to service provider $I$. 
       \item $GenerateChallenge(PK_I) \rightarrow c$: This algorithm generates a random challenge value, based on the given public key. 
       \item $GenerateKeys(PK, MSK, c, I, S) \rightarrow k$:
       This algorithm performs the key generation. The DO provides the public parameters $PK$, their master secret key $MSK$, the DID of the service provider $ I $, the challenge value $c$, and an optional set of attributes $ S$. The algorithm returns a cryptographic key $k$.
   \end{enumerate}

   \item $Encrypt(PK, A, m) \rightarrow C$: This algorithm uses the public parameters $PK$, the access policy $A$ and the plaintext message $m$, to produce the cipher text $C$.
   
   \item $Decrypt(PK, CT, k) \rightarrow m$ or $\bot$: This algorithm has two steps:
   
   \begin{enumerate}
        \item $ValidateChallenge(k, sk) \rightarrow b$: This step extracts the random challenge value $c$ from the key $k$ , to ensure ownership of the private key $sk$ by the intended service provider.  This step returns $b$ as either True or False. 
        
        \item $Dec(PK, CT, k, b) \rightarrow m$ or $\bot$: This step takes the public parameters $PK$ and the  key $k$ to decrypt the ciphertext $CT$. It returns the message $m$ if $b$ is True and the attributes within the key satisfy the access structure embedded within the ciphertext. 
   \end{enumerate}
\end{itemize}

\subsection{Network model}
.The network process consist of the following major steps involving a user, a service provider and the Indy network: 

\begin{enumerate}
    \item The SP generate a cryptographic keypair and submit its DID to the Indy network. 
    \item The SP's access policy structure is published to the Indy network. The access policy is described as an XACML data structure \cite{lorch2003first}. 
    \item The user navigates to the service provider's online form. 
    \item The user initiates the process of creating one or more DCs intended for the service provider. 
    \item The user engages in a ceremony with the service provider to obtain the specific DIDs of the entities that will consume the DCs. 

\item For each DC, the user generates a key by executing the $KeyGen$ algorithm. The $KeyGen$ expects the DID of the recipient. The algorithm performs a DID lookup to obtain the corresponding public key from the network. A random challenge value is provided and assigned as one of the attributes within the key. The key may contain other attributes.  By using the challenge as an attribute, the user ensures that only the intended recipients are able to decrypt the capsule. 
    \item The user produces one or more verifiable presentation. For example a VP may contain the user's first and last name along with their home address. 
    \item The user extract the access policy from the blockchain in the form of a XACML by performing a network lookup using the service provider's DID. 
    \item The user encrypts the VP and the access policy by executing the $Encrypt$ method. The resulting ciphertext is added to the Data Capsule.
    \item The user securely transfers the DCs to the service provider.
    \item The service provider is able to use the key contained in the DC, along with it's own secret keys to execute the  $Decrypt$ step for each DC to decrypt the data as long as the attributes embedded in the key satisfy the access policy.
    
\end{enumerate}

\section{Related Work}

In paper \cite{9140747} the authors present B-Box, a privacy-preserving distributed storage system based on InterPlanetary File System (IPFS) \cite{benet2014ipfs} and Ethereum blockchain. In this system the documents are encrypted using Multi-Authority ABE, and the hash of documents returned by IPFS is stored in Ethereum blockchain. 
 
Axin Wu et al. \cite{Wu2019} propose a  privacy respecting data sharing scheme based on CP-ABE and blockchain, capable of hiding the access control structure by using the attribute bloom filter. 
 
Nabeil Eltayieb et al. \cite{eltayieb2020blockchain} proposed a blockchain-based attribute-based syncryption scheme for the cloud environment.  In their scheme the user's private keys are specified by a set of attributes. The encrypting party can specify a policy over these attributes to determine  which users are able to decrypt the data. In the initial ABE access scheme proposed in \cite{sahai2005fuzzy}, the ciphertext and the user's key are associated with a set of attributes. In addition, a threshold value $k $ is added to the user's key. A user is able to decrypt a ciphertext as long as there is at least $k$ attributes overlapped  between the attributes in the private key and the ciphertext. However this form of threshold semantics is not very expressive, and limit the design of more generic models. 

The term 'data capsule' has been used in different context and for different applications including privacy preserving data models \cite{wang2019data}\cite{kannan2010data}\cite{8688387}. 
In paper \cite{wang2019data} the authors describe a container of subject's data along with its access policy governing how the data is processed. The authors introduce a formal privacy policy language capable of encoding data privacy requirements.
In paper \cite{kannan2010data}, the authors proposes a model to encapsulate sensitive personal data along with policies that support fine-grained access control, with support for data transformation before data is shared with service providers to ensure data minimization. 
While in paper \cite{8688387}, the authors  define data capsule as generic representational specification of data for cloud-edge computational architecture for autonomous robots. 

The existing body of literature work reviewed demonstrates a variety of approaches to building privacy centric data models, however our proposed model is the first work to leverage SSI principles, DID and VC standards to provide a user-centric data model with support for granular access control based on ABE.

\section{Conclusion and Future Work}

In this paper we proposed a Data Capsule model. A self-contained data model consisting of verifiable data, access policies and cryptographic keys to define and enforce specific access policies.   Our model is designed based on Hyperledger Indy, Decentralized Identifiers,  Verifiable Credentials and attribute-based Encryption.  By requiring the service providers to publish their access policies to the blockchain, our model allows proper oversight of policies, and ensures strict conformance by the service providers. Furthermore, the proposed model eliminates the need for the availability of users or third party data custodians during every data access request.
For future work we intend to publish our model implementation consisting of a test network, simulated actors and a digital wallet capable of exchanging Data Capsules.

\printbibliography

\end{document}